\documentclass{iaus}
\usepackage{graphicx}

\title[IAU 281.~~The Highlights] 
{Binary Paths to Type Ia Supernovae Explosions: The Highlights}

\author[Lilia Ferrario]   
{Lilia Ferrario}

\affiliation{Department of Mathematics\\ Mathematical Sciences Institute\\
The Australian National University, ACT 0200\\ 
Australia\\
email: {\tt Lilia.Ferrario@anu.edu.au}}

\pubyear{2012}
\volume{281}  
\pagerange{xxx}
\jname{Binary paths to type Ia supernovae explosions}
\editors{R. Di Stefano, M. Orio, \& M. Moe, eds.}
\begin{document}

\maketitle

\begin{abstract}
This Symposium was focused on the hunt for the progenitors of
Supernovae of Type Ia. Is there a main channel for the production of
SNeIa? If so, are these elusive progenitors Single Degenerate or
Double Degenerate systems?  Although most participants seemed to
favour the Single Degenerate channel, there was no general agreement
on the type of binary system at play. An observational puzzle that was
highlighted was the apparent paucity of Super-Soft Sources in our
Galaxy and also in external galaxies. The Single Degenerate channel
(and as it was pointed out, quite possibly also the Double Degenerate
channel) requires the binary system to pass through a phase of steady
nuclear burning. However, the observed number of Super-Soft sources
falls short by a factor of up to 100 in explaining the estimated birth
rates of SNeIa. Thus, are these Super-Soft sources somehow hidden away
and radiating at different wavelengths or are we missing some
important pieces of this puzzle that may lead to the elimination of a
certain class of progenitor?  Another unanswered question concerns the
dependence of SNeIa luminosities on the age of their host
galaxy. Several hypotheses were put forward, but none was singled out
as the most likely explanation.

It is fair to say that at the end of the Symposium the definitive
answer to the vexed progenitor question remained is well and truly
wide open.

\end{abstract}

\firstsection 
\section{Introduction}

The long standing problem of establishing which objects produce SNeIa is
often referred to as the ``progenitor problem'' and the main aim of
this Symposium was to shed some light on this decade old question. 

It is widely acknowledged that SNeIa are generated by thermonuclear
disruptions of carbon-oxygen (CO) white dwarfs (WDs). Thus, according
to the most commonly accepted view, a WD must either accrete enough
matter from its companion or it must merge with its companion. Hence
the following two principal channels for the formation of SNeIa:

\begin{enumerate}
\item Single Degenerate (SD) channel: A WD accretes mass from a
  non-degenerate companion star, reaches the Chandrasekhar's limiting
  mass and explodes. This channel was first proposed by
  \cite{Whelan1973}. Alternatively, a WD accumulates a large enough
  mass of helium on top of its CO core either via accretion or through
  the burning of H into He. The helium layer ignites and creates a
  burning front propagating outwards through the helium shell while an
  inward compressional wave ignites the CO causing a detonation which
  propagates outwards (``double-detonation'' model,
  \cite[Woosley \& Weaver 1994]{Woosley1994}).

\item Double Degenerate (DD) channel: Two WDs merge to form a single
  WD achieving the Chandrasekhar's limiting mass which then
  explodes. This pathway was first proposed by \cite{IbenTutokov1987}
  and \cite{Webbink1984}.
\end{enumerate}

The SD channel received a major boost following the discovery of the
Supersoft X-ray Sources (SSSs, e.g. \cite[Tr\"umper
  et al.  1991]{Truemper1991}), which are binary systems consisting of
a WD which accretes hydrogen at a rate that is high enough ($\ge
10^{-7}$ M$_\odot$ per year) to allow steady nuclear burning over a
period of $\sim 10^5-10^6$ years.  Thus, if these WDs can accumulate
$\sim 0.1-0.4$ M$_\odot$ of material during their nuclear burning
phase, they could end up as SNeIa.

In the case of stable Roche Lobe Overflow (RLOF) the donor does not
need to be a hydrogen-rich main sequence (MS) or giant-like star. It
could be a non-degenerate or a semi-degenerate helium-burning star, or
even a helium WD (e.g. \cite[Iben et al.  1987]{Iben1987}).

Another interesting avenue was presented by Soker (this volume). This
consists of a merger between a WD with the hot core of a massive AGB
star during Common Envelope (CE) evolution.

Needless to say, the Symposium was entirely focused on discussions on
the pros and cons of the two main channels (DD versus SD) both from
observational and theoretical points of view.

\section{Single Degenerate channel}\label{SD_channel}

In the SD channel, a massive WD needs to steadily burn hydrogen and
increase its mass to reach the Chandrasekhar's limit.

The symbiotic stars (SSs - see the reviews by Mikolajewska, this
volume) are long orbital period ($\sim 10^2-10^3$ days) interacting
binaries in which a hot WD orbits around a cool M giant or a Mira
variable. They are usually immersed in an emission nebula and there is
no direct evidence for the existence of an accretion disc.

Mass transfer occurs via Bondi-Hoyle-Littleton (BHL) wind accretion,
although the recent detection of ellipsoidal variability in some SSs
with $P_{\rm orb}<10^3$ days suggests that some of the giant donors
are close to filling their Roche lobes. During quiescence, SSs undergo
stable hydrogen burning on the WD. Thermonuclear nova explosion are
rare in classical SSs (9 objects out of $\sim 200$).

Recurrent Novae (RNe, see the review by Anupama, this volume) exhibit
nova explosions with mass ejection of $10^{-7}-10^{-6}$ solar masses
per year on recurrence timescales of $\sim 10- 10^2$ years. RNe can
be divided into two subclasses: (i) systems with a dwarf secondary and
(ii) systems with a red giant secondary. Those with a giant
secondaries, such as RS Oph, T CrB, V3890 Sgr and V745 Sco, bear many
similarities with the symbiotic stars. Thus, these are often referred
to as ``Symbiotic Recurrent Novae'' (SyRNe). Recent observations of
ellipsoidal variations in all SyRNe, except for RS Oph, indicate that
mass accretion in most SyRNe occurs via Roche Lobe Overflow (RLOF).

Interestingly, the theoretical calculations of Mohamed \&
Podsiadlowski (this volume) show that standard wind accretion is not
adequate to explain the observations of SSs. Thus, they propose a
model where the wind acceleration region of the evolved star occurs
close to its Roche lobe. Thus, the wind material can fill the Roche
lobe and matter can be transferred to the hot star through the inner
Lagrangian point. The mass transfer rate is estimated to be up to 100
times larger than that expected from BHL wind accretion.

The orbital parameters are known for only two SyRNe: T CrB and RS Oph
(e.g.  \cite[Nelson et al.  2011]{Nelson2011}, \cite[Belczynski \&
  Mikolajewska 1998]{Belczynski1998}). The orbital periods are amongst
the shortest observed for SSs: 227 and 453 days, respectively. The
systems T CrB and RS Oph are also the only SSs with the hot and
massive WD ($1.1-1.4$ M$_\odot$) being more massive than the cool
giant companion whose mass is only $\sim 0.6$ M$_\odot$ (much lower
than that of any other symbiotic giant).  Mukai et al. (this
volume) and Orio, Nelson \& Rauch (this volume), show how X-ray
observations of accreting WDs can be used to set lower limits to the
mass, temperature and chemical composition of WDs in RNe.

Thus, observations indicate that the masses of the WDs in a fraction
of SyRNe are sufficiently high to allow them to become SNeIa if enough
material is accreted (see Kato, this volume).  The low mass of the
nova ejecta and the absence of WD processed material indicate that the
WDs in these systems burn steadily the accreted material and during
eruption they do not expel more matter than they have accreted. Hence,
the mass of the WD can grow and SyRNe could in principle provide a
good avenue for the production of SNeIa (see also the calculations of
Starrfield et al., this volume).

However, it is not at all clear whether these WDs were either (i)
relatively low-mass CO WDs that have accreted a significant amount of
mass during binary evolution or (ii) they were massive ONeMg WDs at
birth. In the first case (i) a SNIa event becomes possible. In the
second case (ii), these ONeMg WDs will undergo an accretion induced
collapse (AIC) and become a neutron star (\cite[Yoon, Podsiadlowski \&
  Rosswog 2007]{Yoon2007}). In fact, population synthesis calculations
have shown that this route can provide a much better match to the
period distributions of long period binary Millisecond Radio-Pulsars
(MSPs) than those that result from core-collapse SNeII and
subsequently go through a Low-Mass X-ray Binary phase (\cite[Hurley
  et al. 2010]{Hurley2010}).

Regardless of the final outcome, massive WDs accreting and burning
hydrogen at high rates [i.e. steady nuclear burning WDs, (NBWDs)] are
hot, bright ($L\sim 10^{38}$ erg s$^{-1}$) and should be observable as
SSSs. However, only $\sim 380$ SSSs have been discovered in external
galaxies which is a number that is up to two orders of magnitudes too
small to match the observed SNeIa rate (Di Stefano, this
volume). Thus, if the observed population of SSSs is so much smaller
than required, we are led to hypothesise that either (i) the SD
channel is just a very minor route to the production of SNeIa; or (ii)
most of the time these NBWDs do not look like SSSs (Di Stefano, this
volume). In the context of (ii), Nielsen, Dominik \& Nelemans (this
volume) have built a model where obscuration by gas within the system
hides most of the nuclear burning phase of these WDs making them
practically unobservable to current X-ray Telescopes. But then we
should try to answer the following question: ``If the energy generated
by steady nuclear burning does not emerge at X-ray wavelengths, in
which wave-bands should these sources be looked for?''

Wang \& Han (this volume) used Eggleton's stellar evolution code with
the optically thick wind assumption to study the SD channel where the
donor is a helium star (either main-sequence or a He sub-giant). They
found that this channel can produce the class of SNeIa with short
delay times ($<100$ Myr). They also predict that the surviving
companion would be characterised by a spatial velocity of $\ge 400$
km/s (see \S \ref{runaway}).

\subsection{Super-Chandrasekhar's SNeIa}

Recent observations of a class of super-luminous SNeIa indicate that
some exploding WDs may have masses of up to $2.4-2.8$ M$_{\odot}$,
well above the Chandrasekhar mass limit (e.g. \cite[Howell
  et al.  2006]{Howell2006}). One of these objects is SNIa 2009dc whose
late-phase observations were reported by Yamanaka et al.  (this
volume).  Hachisu et al.  (this volume) presented a binary scenario
where accreting WDs can reach masses as high as $2.4$ M$_\odot$
supported by rotation against collapse and explosion. Magnetic-dipole
radiation would lead to spin-down on time-scales of $<10^9$ years in
WDs with $M\ge 1.6$ M$_\odot$ and $>10^9$ years in WDs with
$1.38<M/M_\odot<1.5$. The range in spin-down time-scales would thus
produce both the prompt and the delayed variety of SNeIa (see
\S \ref{DTD}).

This scenario is similar to that recently reported by
\cite{Distefano2011}. Their model predicts that by the time a SNIa
explosion occurs, the donor star will have had sufficient time to lose
its envelope and perhaps become a WD. This would have the benefit of
reducing the impact that a hydrogen-rich donor would have on the SN
characteristic signatures.  Before the explosion these systems could
appear either as super-Chandrasekhar mass WDs with a WD companion or as
CVs.  After the explosion, the model forecasts the existence of a
population of either WDs, or low-mass stars with high spatial
velocities. This scenario would also have the added benefit of
explaining the existence of single low-mass WDs (M$<0.5$ M$_\odot$) as
the remnants of giant-branch donors whose envelopes were stripped away
by the supernova explosion (\cite[Justham et al.  2009]{Justham2009}).

\subsection{Runaway Donor Stars in the SD Channel}\label{runaway}

The SD route to SNeIa predicts that the donor star should survive a
SNIa explosion (e.g. \cite[Marietta, Burrows \& Fryxell
  2000]{Marietta2000}). Consequently, if such star could be found,
this would provide substantial evidence in support of this channel.

Kerzendorf et al. (this volume) have argued that a left over
donor star could be identified by its space velocity. Furthermore, if
one assume RLOF, the donor star is expected to be tidally locked to the
rotation period of the system at the time of explosion and thus be
rapidly rotating.

They analysed 79 stars near the centre of the remnant of SN~1006 (at
the distance of 2.2 kpc), but found no stars, down to half a solar
luminosity, displaying an unusual space or significant rotational
velocities (down to $< 20$ km s$^{-1}$). They also re-examined a
sub-giant star (Star~G) which was identified by \cite{Lapuente2004} as
the donor star of Tycho's 1572 SNIa. Kerzendorf et al. argued
that although Star~G exhibits an unusual kinematic, it is off-centre
and shows no rotation. Furthermore, they could not confirm the
presence of Nickel in its atmosphere as reported by
\cite{Hernandez2009}. Thus, they argue that Star~G is not likely to be
associated with the SN event.

Ruiz-Lapuente et al. (this volume) also conducted a survey down
to $R = 15$ of the stars within a 4 arcmin radius around the remnant
of SN~1006 for a possible surviving binary donor. The limiting
magnitude of the survey would have allowed the detection of red-giant
type stars, but their results were also inconclusive.

\cite{Podsiadlowski2003} reports that to exclude a sub-giant companion,
it is necessary to consider much fainter objects than previously
forecast. For example, one should look for stars with a luminosity of
0.1 solar luminosities in a 1,000 year old remnant, or 0.01 solar
luminosities in a 10,000 year old remnant.  Thus, deeper surveys of
SNIa remnants may be needed to look for these runaway donor stars.

However, if these deeper surveys still cannot identify a suitable
candidate, this would indicate that other progenitor scenarios, such
as the DD channel (see \S \ref{DD_channel}), are at work or
that there is a long enough time delay between the accretion phase and
the SNIa explosion that would allow the donor to become a He WD, as
expected in some models (e.g. \cite[Di Stefano, Voss \& Claeys
  2011]{Distefano2011}).

\subsection{Classical CVs and WD Masses: A Case for Failed SNeIa?}

Theoretical simulations and observations indicate the WD mass in CVs
decreases over time due to nova eruptions. However,
\cite{Zorotovic2011} have discovered that recent high-precision
measurements of WD masses in a large number of CVs point to primary
masses which are significantly higher ($0.8-0.9$ M$_\odot$ per year)
than the average mass of isolated WDs or of WDs in pre-CVs ($\sim 0.6$
solar masses). Taken at face value, these observations may indicate
that either (i) the amount of material ejected during nova eruptions
in classical CVs is less than what is accreted and thus the WD mass
increases during CV evolution or (ii) most CVs have undergone an early
SSS phase during which a mass increase occurred.

In case (i), if the WDs can accrete enough mass to trigger an
explosion, these could become SNeIa with long delay times.  In case
(ii), the WDs would have already reached their maximum mass at the end
of their SSS phase. These CVs could thus be seen as ``failed
SNeIa''. This view seems to be supported by far-UV observations of CVs
showing that about 10\% of systems exhibit a large NV/CIV emission
line flux ratio, probably due to accretion of CNO-processed material
(\cite{deMartino2009}).

\section{The Double Degenerate Channel}\label{DD_channel}

According to the DD channel, two WDs merge and reach the
Chandrasekhar's limiting mass. The less massive of the two WDs is
tidally disrupted and forms an accretion disc around its more massive
companion. Carbon is then expected to ignite at the core-envelope
boundary of the accreting WD with the nuclear flame propagating
inwards (e.g. \cite[Nomoto \& Iben 1985]{Nomoto1985}). This would turn
the CO WD into an ONeMg WD. If the mass of the ONeMg WD reaches the
Chandrasekhar's limit, electron capture on Ne and Mg will cause an AIC
of the WD into a neutron star. This seems to be the nearly inevitable
fate of the merger of two WDs and has been the greatest stumbling
block for this channel to be considered as a viable route to SNeIa.

However, under very specific conditions, AIC may be avoided.  If
carbon ignites in the deep interior of the WD and is triggered in a
way to switch from subsonic deflagration to supersonic detonation at
the correct time, then one would obtain the observed explosion
characteristics and the production of the correct mixture of
elements.

Zhu, Chang \& van Kerkwijk (this volume) have explored how much fine
tuning is necessary to produce the suitable conditions that lead to
SNeIa for both Chandrasekhar and sub-Chandrasekhar mass CO-CO WD
mergers and thus describe how central temperature, density, and merger
remnant morphology critically depend on mass of the component stars
and the mass ratio. They find that if mergers occurs between roughly
equal-mass CO WDs, the merger remnants are fully mixed, hottest at the
centre and surrounded by dense and small accretion discs. If the disc
is accreted quickly, the compressional heating would likely lead to
central carbon ignition at densities at which a detonation would lead
to SNeIa. With this merger scenario, the SNeIa rates would be matched,
since lower mass WDs could participate in the production of SNeIa and
would explain the observed range of luminosities and the fact that the
bright SNeIa are found in younger stellar populations (see also Sim
et al. this volume).

Interestingly, Di Stefano has pointed out that most systems that
evolve into DDs must also undergo a long-lived ($>10^6$ years) NBWD
phase as SSSs and thus are likely to spend a fraction of their lives
as symbiotic systems.  Hence we may still have to deal with the
possibility that NBWDs may not appear as SSSs most of the time (Di
Stefano, this volume), in order to reconcile their paucity with the
existence of their progenies.

\subsection{DD Merger on a Dynamical Timescale}

Webbink pointed out that among double WDs massive enough to be SNIa
candidates, the mass transfer process is nearly always dynamically
unstable.  That instability tidally disrupts the donor (less massive)
WD within a few orbits, i.e., within a few minutes.  Effective
accretion rates can reach $10^5$ M$_\odot$ per year, far in excess of
the nominal Eddington limit, so fast that photons have no time to
diffuse, but are advected along with the gas flow.  Much of the
initial orbital angular momentum of the binary is carried off in tidal
tails shed during disruption of the donor, and its disrupted remnant
forms a rapidly rotating (but predominantly pressure-supported)
envelope around the accreting WD.  Shock heating may lead to carbon
ignition at the base of this envelope, and, provided that burning is
not quenched by expansion, it will percolate inward on a roughly
thermal time scale.  However, this process may be completely preempted
by adiabatic compression of the accretor core, leading to immediate
central CO ignition, triggering a SNIa event. The calculations of
Scannapieco, Raskin \& Timmes (this volume) based on adaptive mesh
refinement simulations of DD mergers and collisions seem to (broadly
speaking) support this scenario.

Crucial questions that need to be addressed are (i) the degree of
internal heating by tidal dissipation in both accreting and donor WDs
preceding merger; (ii) the extent to which degeneracy of the donor may
be lifted by accretion shocks during the merger process; (iii) what
becomes of the remaining angular momentum of the merged WDs; and (iv)
under what circumstances does compression of the accretor succeed or
fail to lead to central carbon ignition.  Detailed calculations are
clearly needed to answer these questions.

\subsection{Surveys of Double WDs}

At this symposium, Marsh reviewed the observational evidence for
populations of both detached and semi-detached DDs.  Double WDs whose
total mass is larger than the Chandrasekhar limit and that will merge
through gravitational radiation within a Hubble time have long been
proposed as possible SNeIa progenitors.  However, even though many
surveys have been dedicated to their discoveries,
(e.g. \cite[Bragaglia et al.  1990]{Bragaglia1990}, \cite[Marsh,
  Dhillon \& Duck 1995]{Marsh1995}, \cite[Napiwotzki, Christlieb,
   \& Drechsel 2003]{Napiwotzki2003}) no objects have so far been
unequivocally identified. On the other hand, considering that only one
double WD in one thousand needs to possess these characteristics to
match the observed SNeIa rates (\cite[Nelemans
  et al.  2001]{Nelemans2001}) and since only $\approx 1000$ WDs have
so far been surveyed for close companions, the current lack of secure
candidates is not as yet very significant.

\cite{Kilic2011} have undertaken an MMT survey of all previously
identified extremely low-mass (ELM) WDs ($M\sim 0.2$ M$_\odot$) from
the SDSS DR4 to study their binary fraction and find possible
mergers. They found 12 systems with a combined mass of
$0.4<M/M_\odot<1.4$ that will merge within a Hubble time. The highest
mass system is J$1233+1602$ with a combined mass of $1.37$
M$_\odot$. These systems could evolve into a variety of objects such
as under-luminous SNeIa, extreme helium stars (RCrB), or single
helium-enriched sub-dwarfs. They also remark that sub-Chandrasekhar
explosions may not lead to ``classical'' SNeIa but the number found
seems to be consistent, according to \cite{Brown2011} with the
observed rate of under luminous SNeIa. However, the calculations of
Sim et al. (this volume) show that sub-Chandrasekhar explosions
may in fact represent an important route to SNeIa events (see \S
\ref{subchandra}).

\subsection{Sub-Chandrasekhar's Explosions}\label{subchandra}

SNeIa could arise from detonations of sub-Chandrasekhar WDs, as
reported by Sim et al. (this volume). They presented a
numerical model that is relevant to any class of sub-Chandrasekhar
mass explosion, regardless of what triggers the detonation mechanism.

Sim et al. considered pure detonations of sub-Chandrasekhar CO
WDs with different masses. For $0.97<M/M_\odot<1.15$ they obtained a
$^{56}$Ni yield of $\sim 0.3-0.8 M_\odot$ and were able to reproduce
almost the full range of brightness observed in SNeIa with rise and
decline timescales and peak colors which are in good qualitative
agreement with the observations of SNeIa. The theoretical spectra
corresponding to light maximum exhibit features due to
intermediate-mass elements and match the correlation between the
explosion luminosity and the ratio of the SiII lines at $\lambda 6355$
and $\lambda 5972$.

Sub-Chandrasekhar models have the benefit that, because of the lower
densities of these WDs, a detonation does not burn the entire WD into
iron-peak elements. That is, these models do not need a finely tuned
switch between deflagration and detonation. On the other hand, if the
triggering mechanism is due to the ignition and detonation of a layer
of helium, which would drive a shock propagating into the WD interior,
then we are left with the following problem. A $\sim 0.2$ M$_\odot$
layer of helium would burn to $^{56}$Ni, giving early-time spectra
rich in $^{56}$Ni, which are not observed. However,
\cite{Bildsten2007} have shown that a detonation can occur in a helium
layer of just $0.06-0.02$ M$_\odot$ on $1.0-1.36$ M$_\odot$ CO
WDs. Although their calculations did not follow the shock into the WD
core, they speculated that it may drive a CO detonation in or near the
stellar centre. \cite{Guillochon2010} have also recently presented a
new mechanism for the detonation of a sub-Chandrasekhar CO WD in a
dynamically unstable system.

Sim et al. (this volume) point out that these could in fact be
an important channel to SNeIa, since the mergers of low-mass WDs would
also come into play. These would yield much higher rates than those
derived from Chandrasekhar mergers, as required by observations of
SNeIa birthrates.

\section{Core-Degenerate (CD) Channel}

Soker (this volume) proposed a new mechanism for the production of
SNeIa which consists of a merger, occurring at the end of the CE
phase, between a WD and the degenerate core of its massive asymptotic
giant branch companion star ($5-8$ M$_{\odot}$). This merger would
occur while the core is still hot and relatively large and would form
a centrifugally supported rapidly rotating WD with a mass that could
exceed the Chandrasekhar's mass limit. On this model, the delay from
stellar formation to explosion would be determined by the spin-down
time of the rapidly rotating merger remnant. Gravitational radiation
was shown to be inefficient in spinning down the WD, while spin-down
time-scales for magnetic-dipole radiation was shown to be consistent
with the delay times required to explain SNeIa.

Population synthesis calculations are needed to investigate whether
the number of such systems is consistent with the birthrate of
SNeIa. Also, it is not completely clear whether the merged remnant, as
it spins down and collapses, would instead yield a long $\gamma$-ray
burst and a rapidly spinning magnetar (e.g. \cite[Tout
  et al.  2011]{Tout2011}). Long $\gamma$-ray bursts are usually
associated to type Ib/c SNe.

\section{Delay Time Detonation and Population Studies}\label{DTD}

The Delay Time Detonation (DTD) of SNeIa is the rate at which SNeIa
occur after a stellar formation burst. Clearly, different progenitor
scenarios yield different delay times between the birth of the binary
system and the SNIa event.  In the SD channel, the time elapsed from
the formation of the binary and the SNIa explosion is given by the
donor mass and thus by its main sequence lifetime. Under the DD
scenario, in a younger universe the only WDs that could merge were the
most massive ones originating from more massive
progenitors. Consequently, a study of DTDs provides an excellent
method to investigate all possible progenitor channels.

Recent observations of SNeIa indicate that their physical parameters
depend on the type of host galaxy.  SNeIa are more common (by at least
one order of magnitude per unit stellar mass), intrinsically more
luminous and with wider light-curves in late-type galaxies with
ongoing star-formation than they are in elliptical galaxies
(e.g. \cite [Mannucci et al.  2005]{Mannucci2005}). Sub-luminous SNeIa
are found mainly in galaxies with older stellar populations. They also
appear to occur exclusively in massive galaxies. This is in stark
contrast to ``normal'' SNeIa which occur across more diverse types of
hosts. As a consequence, it is to be expected that brighter and wider
light-curve SNeIa should be more common in the distant universe than
they are in the local universe (\cite[Howell
  et al. 2007]{Howell2007}).  Thus, all the available observational
evidence seems to indicate that the local SNIa samples are different
from the high redshift ones.  These redshift-related discrepancies are
of concern to cosmological studies, since they depend on SNeIa being
reliable ``standard candles''. It is therefore important to establish
whether the same light curve ``stretching'' that is calibrated and
used at low redshfits is also applicable at high redshifits (see
\cite[Greggio 2010]{Greggio2010}).

\cite{Scannapieco2005}, \cite{Mannucci2006} and \cite{Sullivan2006}
suggested that these effects may point to the coexistence of more than
one class of SNIa progenitors. In particular, they showed that
delay-time distributions with both ``prompt'' ($100-500$ Myr) and
``delayed'' ($\sim 5$ Gyr) components, or with a wide range of
delay-times, fit the observations quite well. Pritchet reported at this
symposium a DTD with a steeper power-law slope and
found differences in the DTD as a function of light curve
characteristics, which also may point to the coexistence of more than
one class of SNIa progenitors.

In this context, the studies of \cite{Townsley2009} are very
interesting, since they show how the metallicity and the average age
of the parent stellar population can influence the characteristics of
SNeIa. In particular, they find (in agreement with observations) that
the average luminosity of SNeIa should decrease with stellar
population age, since colder WDs have higher core densities and
produce less $^{56}$Ni (possibly due to a higher rate of
neutronisation).

In summary, the importance of the analysis of the DTDs at higher
redshifts cannot be emphasised enough, since it will help us
understand how the DTD evolves over time. To this purpose, Reg\"os et
al. (this volume) are planning a survey to investigate SNeIa at
redshfift up to $z\sim 2$.

\section{Conclusions}

Almost 80\% of SNeIa are ``normal'' in the sense that they are
virtually carbon copies of each other, while the remaining 20\% appear
quite different and can be described as ``extremely weak'', ``weak''
or ``extremely luminous''. The spread in observed peak luminosities,
differences in light curve characteristics, delay time distributions,
and the dependence of physical parameters on the type of host galaxy,
all appear to suggest that more than one channel may be responsible
for SNeIa.

Discussions of the ``normal'' SNeIa were focused on the explosion
properties of Chandrasekhar mass WDs where a WD is driven to the
Chandrasekhar's limit by accretion in a binary system. There was no
consensus on the nature of the binary star progenitor, although most
participants appeared to favour the SD hypothesis. The SD hypothesis
requires the system to go through a phase of stable nuclear burning on
the surface of the WD, but the known number of SSSs leads to estimates
for birth rates of SNe Ia that fall short of the required birth rate
by several orders of magnitude. Either the majority of NBWDs radiate
at other wavelengths, or SD progenitors only contribute a small
fraction of the total SNeIa rate. This, together with the lack of
evidence of hydrogen in the spectra of these systems were discussed as
major problems for the model from an observational point of view. A
criticism of the SD model was that it had a ``lot of moving parts''
and could only explain the observed delay time $\propto t^{-1}$ in a
``contrived way'', while this relationship was explained more
naturally in the DD models (see the calculations of \cite[Ruiter,
  Belczynski \& Fryer 2009]{Ruiter2009}).

As has been known for sometime, all Chandrasekhar explosion models
require an ad hoc transition between the time of deflagration
and detonation to generate light and abundance curves that agree with
observations.  None of the papers presented at the conference yielded
new theoretical insights on this problem.  For the DD channel to work
there was the additional constraint that the accreting CO WD should
detonate at the centre otherwise it would be transformed into an ONeMg
WD which would then collapse into a neutron stars via an AIC.  Again,
there were no calculations presented at the conference that directly
addressed this problem although there was some discussion led by
Webbink around the idea that in the DD channel compressional heating
of the core caused by very high accretion rates from a dynamically
unstable accretion disc may play a dominant role in triggering central
detonation.

A highlight of the conference was perhaps the demonstration that
detonation models could be constructed for WDs with masses ranging
from sub-Chandrasekhar to Chandrasekhar. These could explain the
entire range of SNeIa light and spectral characteristics from
extremely weak through to normal SNeIa, opening up a variety of
possibilities for DD models. However, this model also suffers from the
lack of an easily identifiable triggering mechanism.

All in all, it is fair to say that despite the significant progress
made on both observational and theoretical fronts, the nature of the
progenitors of SNeIa still remains unresolved.

\end{document}